\documentclass[conference]{IEEEtran}
\IEEEoverridecommandlockouts
\usepackage{amsmath,amssymb,amsthm}
\usepackage{enumerate}
\usepackage{stfloats}
\usepackage{comment}
\usepackage{bbm}
\usepackage{mathtools}
\usepackage{accents,latexsym,cancel}
\usepackage{cite}
\usepackage[nolist]{acronym}

\usepackage[dvipsnames]{xcolor}
\definecolor{myPink}{RGB}{255,105,183}

\usepackage[T1]{fontenc}
\usepackage{graphics} % for pdf, bitmapped graphics files
\usepackage{epsfig} % for postscript graphics files
\usepackage[mathscr]{euscript}
\usepackage{algorithm}
\usepackage[noend]{algpseudocode}
\usepackage{bbm}
\makeatletter
\def\BState{\State\hskip-\ALG@thistlm}
\makeatother

\usepackage{tikz}
\usetikzlibrary{arrows,shapes,chains,matrix,positioning,scopes,patterns,calc,
decorations.markings,
decorations.pathmorphing,
}

\usepackage{pgfplots}
\pgfplotsset{compat=1.3}
\usepgflibrary{shapes}
\usetikzlibrary{pgfplots.fillbetween}

\newtheorem{theorem}{Theorem}

\newtheorem{remark}[theorem]{Remark}

\renewcommand{\epsilon}{\varepsilon}

\DeclareMathOperator*{\prox}{prox}

\newcommand{\RNum}[1]{\uppercase\expandafter{\romannumeral #1\relax}}

\newcommand{\rv}{\ensuremath{\mathbf{r}}}
\newcommand{\sv}{\ensuremath{\mathbf{s}}}

\newcommand{\vv}{\ensuremath{\mathbf{v}}}

\newcommand{\wv}{\ensuremath{\mathbf{w}}}

\newcommand{\yv}{\ensuremath{\mathbf{y}}}
\newcommand{\zv}{\ensuremath{\mathbf{z}}}

\newcommand{\betav}{\ensuremath{\boldsymbol{\beta}}}

\newcommand{\Xm}{\ensuremath{\mathbf{X}}}

\DeclareMathAlphabet{\mcl}{OMS}{cmsy}{m}{n}

\DeclareMathOperator*{\argmin}{\,arg\ min}

\newlength\tikzwidth
\newlength\tikzheight

\definecolor{mycolor1}{rgb}{0.63529,0.07843,0.18431}%
\definecolor{mycolor2}{rgb}{0.00000,0.44706,0.74118}%
\definecolor{mycolor3}{rgb}{0.00000,0.49804,0.00000}%
\definecolor{mycolor4}{rgb}{0.87059,0.49020,0.00000}%
\definecolor{mycolor5}{rgb}{0.00000,0.44700,0.74100}%
\definecolor{mycolor6}{rgb}{0.74902,0.00000,0.74902}%
{\begin{list}{$\bullet$}
   {\setlength{\itemsep}{0in}
    \setlength{\labelsep}{0.05in}
    \setlength{\labelwidth}{0.2in}
    \setlength{\leftmargin}{0.25in}
    \setlength{\rightmargin}{0in}
    \setlength{\topsep}{0in}
   }}
{\end{list}}

\begin{document}

\title{Approximate Message Passing for Multi-Preamble Detection in OTFS Random Access}

\author{%
  \IEEEauthorblockN{Alessandro Mirri$^1$, Vishnu Teja Kunde$^2$, Enrico Paolini$^1$, Jean-Francois Chamberland$^2$}
  
  \IEEEauthorblockA{$^1$ CNIT/WiLab, Department of Electrical, Electronic, and Information Engineering, University of Bologna, Italy \\
  Email: \{alessandro.mirri7, e.paolini\}@unibo.it}
  
  \IEEEauthorblockA{$^2$ Department of Electrical and Computer Engineering, Texas A\&M University, USA \\
  Email: \{vishnukunde, chmbrlnd\}@tamu.edu}
  
  \thanks{The University of Bologna team is supported by the European Union -- Next Generation EU under the Italian National Recovery and Resilience Plan (NRRP), partnership on ``Telecommunications of the Future'' (PE00000001 -- program ``RESTART'').  
  This material is based, in part, upon work conducted at Texas A\&M University and supported by the National Science Foundation (NSF) under Grant CNS-2148354.
  
  \textbf{This is the author’s preprint version of a work submitted to the IEEE for possible publication. Copyright may be transferred without notice, after which this version may no longer be accessible on arXiv.}}
}

\maketitle

\begin{acronym}
% usage: \ac{SW}, \acp{SW} for plurals \acf{SW} Use the full name of the acronym.
%\acs{SW}Use the acronym, even before the first corresponding \ac command
%\acl{acronym}Expand the acronym without using the acronym itself.
\small
\acro{AMP}{approximate message passing}
\acro{AWGN}{additive white Gaussian noise}
\acro{BS}{base station}
\acro{CAMP}{complex AMP}
\acro{DD}{delay-Doppler}
%\acro{G-LASSO}{Group LASSO}
\acro{GL}{group LASSO}
\acro{SG}{sparse Group}
%\acro{SG-LASSO}{Sparse Group LASSO}
\acro{SGL}{sparse group LASSO}
\acro{CSG}{complex sparse group}
\acro{CSGL}{complex sparse group LASSO}
\acro{CGL}{complex group LASSO}
\acro{CL}{complex LASSO}
\acro{i.i.d.}{independent and identically distributed}
\acro{ISTA}{iterative soft-thresholding}
\acro{LASSO}{least absolute shrinkage and selection operator}
\acro{OFDM}{orthogonal frequency-division multiplexing}
\acro{OMP}{orthogonal matching pursuit}
\acro{OST}{one step thresholding}
\acro{OTFS}{orthogonal time frequency space}
\acro{PHY}{physical}
\acro{SNR}{signal-to-noise ratio}
\acro{Veh-A}{Vehicular A}
\acro{ZC}{Zadoff-Chu}
\end{acronym}

\begin{abstract}
This article addresses the problem of multiple preamble detection in random access systems based on orthogonal time frequency space (OTFS) signaling.
This challenge is formulated as a structured sparse recovery problem in the complex domain.
To tackle it, the authors propose a new approximate message passing (AMP) algorithm that enforces double sparsity: the sparse selection of preambles and the inherent sparsity of OTFS signals in the delay–Doppler domain.
From an algorithmic standpoint, the non-separable complex sparsity constraint necessitates a careful derivation and leads to the design of a novel AMP denoiser.
Simulation results demonstrate that the proposed method achieves robust detection performance and delivers significant gains over state-of-the-art techniques.
\end{abstract}


\section{Introduction}
Many high-dimensional communication problems exhibit sparsity, either due to the physical characteristics of wireless channels or as a result of coding schemes \cite{duarte2011structured,willett2013sparsity,qin2018sparse}.
For example, sparse regression coding achieves optimal asymptotic performance \cite{barbier2017approximate,rush2021capacity}, while coded compressed sensing formulates unsourced random access as a sequence of sparse recovery tasks~\cite{vem2019user}.
Multipath fading also induces sparsity, motivating the use of \ac{OFDM} and, more recently, \ac{OTFS} signaling \cite{hadani2017orthogonal}, which exploits sparsity in the \ac{DD} domain.

Sparsity-inducing methods such as the \ac{LASSO} \cite{tibshirani1996regression} and its algorithmic variants \cite{beck2009fast,cai2011orthogonal}, promote element-wise sparsity but do not account for hierarchical structures.
To address this limitation, structured approaches like \ac{GL} \cite{yuan2006model} and \ac{SGL} \cite{simon2013sparse}
enforce both group and intra-group sparsity, providing a better fit for problems with double sparsity, such as multiple preamble detection in \ac{OTFS}-based random access {\cite{Mirri2025:ZakOTFS_CRA}.
In this setting, only a few preambles are active, and each active signal propagates through a sparse multipath channel.

In parallel, \ac{AMP} algorithms \cite{donoho2009message} have emerged as efficient, theoretically grounded solutions for large-scale sparse recovery.
Extensions to the complex domain, such as \ac{CAMP} \cite{Maleki2013CAMP}, are particularly relevant in wireless applications.
Recently, \ac{AMP} has been extended to incorporate structured sparsity, paving the way for powerful new frameworks.

The main contributions of our article are:
(\textit{i}) an \ac{AMP} algorithm tailored to complex-valued \ac{SGL}, specifically designed for multiple preamble detection in \ac{OTFS}-based random access;
(\textit{ii}) a structured denoiser that jointly exploits group and intra-group sparsity, with a rigorously derived Onsager correction term to enhance convergence; and
(\textit{iii}) extensive simulations demonstrating significant performance gains over state-of-the-art methods.

\section{Background and Problem Formulation}
\label{sec:Background}

This section offers an overview of sparse regression, focusing on structured sparsity in complex-valued settings.
We start with the canonical linear inverse problem
\begin{align} 
\label{equation:ReceivedSignalRaw}
\yv = \Xm \betav + \wv,
\end{align}
where $\yv \in \mathbb{C}^n$ is the received vector, $\Xm \in \mathbb{C}^{n \times N}$ represents the sensing matrix, and $\betav \in \mathbb{C}^N$ is the unknown sparse vector.
Additive noise vector $\wv \in \mathbb{C}^n$ has \ac{i.i.d.} entries, each with complex Gaussian distribution $w_i \sim \mathcal{CN}(0, \sigma_\mathrm{n}^2)$, where $\sigma_\mathrm{n}^2$ denotes the noise variance.
For the problem we wish to study, \eqref{equation:ReceivedSignalRaw} arises from an \ac{OTFS}-based random access scenario.
We analyze this setting in detail, as it shapes both the sensing matrix and the sparsity exploited in recovery.

\subsection{OTFS-Based Random Access Model}
\label{subsec:ZakOTFSModel}

\ac{OTFS} is a two-dimensional modulation scheme designed for robust transmission over high-mobility, doubly-dispersive channels, mapping signals onto a \ac{DD} grid to exploit channel sparsity \cite{SaifBook}.
We consider an uplink random access scenario with $K$ active users, each selecting a preamble $\sv \in \mathbb{C}^{M_\mathrm{dd}N_\mathrm{dd}}$ from a common pool of $G$ sequences, where $M_\mathrm{dd}$ and $N_\mathrm{dd}$ denote the number of delay and Doppler bins, respectively.
Every active user $k \in \{1, \dots, K\}$ randomly selects an index $i_k \in \{1, \dots, G\}$ and transmits the corresponding \ac{OTFS}-modulated preamble $\sv_{i_k}$.
As is typical in random access, only a small fraction of preambles are active at any given time, with $K \ll G$.
The wireless channel between each user and the receiver is modeled as a sparse multipath channel with $L$ dominant \ac{DD} components.
Each multipath component is described by its delay $\tau_{k,l}$, Doppler shift $\nu_{k,l}$, and complex gain $h_{k,l}$, for $l = 1, \dots, L$.
The received \ac{OTFS} signal at the access point is
\begin{equation} \label{eq:ZakOTFSReceived}
\textstyle
\yv = \sum_{k=1}^K \sum_{l=1}^{L} h_{k,l}\,\sv_{i_k}^{(\tau_{k,l}, \nu_{k,l})} + \wv,
\end{equation}
where $\sv_{i_k}^{(\tau_{k,l}, \nu_{k,l})}$ denotes the preamble $\sv_{i_k}$ shifted in the \ac{DD} domain by $(\tau_{k,l}, \nu_{k,l})$, and $\wv$ denotes additive complex Gaussian noise.
This gives rise to a structured sparse recovery problem: a small number of active preambles, each experiencing sparse multipath propagation.

Interpreting \eqref{equation:ReceivedSignalRaw} in this context, we see that $\Xm$ is a structured sensing matrix whose columns correspond to \ac{OTFS}-modulated preambles shifted over discrete sets of delays $\mathcal{T}$ and Dopplers $\mathcal{D}$, defined as
\begin{align}
\label{eq:DDtaushifts}
\mathcal{T} &= \left\{ 0, \frac{\tau_p}{M_\mathrm{dd}}, \dots, \left\lfloor \frac{\tau_{\max} M_\mathrm{dd}}{\tau_p} \right\rfloor \frac{\tau_p}{M_\mathrm{dd}} \right \}, \\
\label{eq:DDnushifts}
\mathcal{D} &= \left\{\!-\!\left\lfloor \frac{\nu_{\max} N_\mathrm{dd}}{\nu_p} \right\rfloor \frac{\nu_p}{N_\mathrm{dd}}, \dots, 0, \dots, \left\lfloor \frac{\nu_{\max} N_\mathrm{dd}}{\nu_p} \right\rfloor \frac{\nu_p}{N_\mathrm{dd}} \right\}.
\end{align}
Above, $\tau_p$ is the delay period, $\nu_p = 1/\tau_p$ is the Doppler period, and $(\tau_{\max}, \nu_{\max})$ denote the maximum delay and Doppler supported by the system.
The Cartesian product $\mathcal{S} = \mathcal{T} \times \mathcal{D}$ defines the full grid of DD shifts used to construct the columns of $\Xm$.
For each preamble $\sv_j$, and for each DD shift $(\tau_i, \nu_i) \in \mathcal{S}$, the corresponding column of $\Xm$ is given by
\begin{align}
\label{eq:ColumnConstruction}
\Xm_{j|\mathcal{S}| + i} = h_{\text{eff},i} \ast_\sigma \sv_j,
\end{align}
where $h_{\text{eff},i}$ denotes the effective channel impulse response associated with the DD shift $(\tau_i, \nu_i)$, as defined in~\cite[Eq.~(13)]{mattu2024delay}, and $\ast_\sigma$ represents the twisted convolution operator.
The unknown vector $\betav \in \mathbb{C}^{G |\mathcal{S}|}$ captures both user activity and channel properties, with two levels of sparsity: group sparsity across preambles and within-group sparsity across DD components.

\subsection{Sparse Regression and Proximal Methods}
\label{subsec:SparseRegression}

Sparse regression aims to recover $\betav$ under sparsity constraints. The original problem
\begin{align}
\label{eq:L0}
\operatorname{min}_{\betav} \| \yv - \Xm \betav \|_2^2 \quad \text{s.t.} \quad \| \betav \|_0 \leq K,
\end{align}
is combinatorial and NP-hard. A common relaxation replaces $\ell_0$ with the convex $\ell_1$ norm, yielding the \ac{LASSO} formulation
\begin{equation}
\label{eq:L1Relaxed}
\operatorname{min}_{\betav} \| \yv - \Xm \betav \|_2^2 + \lambda \| \betav \|_1 .
\end{equation}
This multi-objective optimization balances fidelity and sparsity while remaining amenable to proximal methods.
The proximal operator of the $\ell_1$ norm is
\begin{equation}
\label{eq:prox_L1}
\prox_{\lambda \|\cdot\|_1}(\betav) = \argmin_{\vv} \frac{1}{2} \| \vv - \betav \|_2^2 + \lambda \| \vv \|_1,
\end{equation}
which reduces to elementwise soft-thresholding for real entries
\begin{equation}
\label{eq:SoftThresholding}
\operatorname{prox}_{\lambda | \cdot |}(\beta) =
\begin{cases}
\beta - \lambda & \beta > \lambda \\
0 & |\beta| \le \lambda \\
\beta + \lambda & \beta < -\lambda .
\end{cases}
\end{equation}
For complex vectors, the $\ell_1$ norm becomes
\begin{equation}
\label{eq:ComplexL1}
\textstyle
\| \vv \|_1 = \sum_i |v_i| = \sum_i \sqrt{ \Re(v_i)^2 + \Im(v_i)^2 },
\end{equation}
which is separable across coordinates.
The proximal operator for the complex case become
\begin{align}
\label{equation:ComplexSoftThresholding}
\operatorname{prox}_{\lambda \| \cdot \|_1} (v) =
\begin{cases}
\left( 1 - \frac{\lambda}{|v|} \right) v & |v| > \lambda \\
0 & |v| \le \lambda,
\end{cases}
\end{align}
which preserves the phase of the argument while shrinking its magnitude.
This generalizes the real-valued operator of \eqref{eq:SoftThresholding}.

\subsection{Structured Sparsity: Group and Sparse Group LASSO}
\label{subsec:StructuredSparsity}

The standard LASSO promotes sparsity at the level of individual coefficients but disregards structural relationships within the vector.
In many applications, however, the nonzero components of $\betav$ naturally cluster into groups (or blocks), motivating the use of \ac{GL}, which enforces group sparsity through
\begin{align}
\label{eq:GroupLASSO}
\textstyle
\min_{\betav} \| \yv - \Xm \betav \|_2^2 + \lambda \sum_{g \in \mathcal{G}} \| \betav_g \|_2,
\end{align}
where $\mathcal{G}$ partitions indices into disjoint groups and regularizers $\| \betav_g \|_2$ encourage low-energy blocks to vanish.
When sparsity occurs both within and across groups, the \ac{SGL} formulation is more appropriate, combining $\ell_1$ norm (within-group sparsity) and $\ell_2$ norm (group sparsity)
\begin{align}
\label{eq:SGL}
\textstyle
\min_{\betav}\| \yv - \Xm \betav \|_2^2 + \lambda_1 \| \betav \|_1 + \lambda_2 \sum_{g \in \mathcal{G}} \| \betav_g \|_2\,,
\end{align}
with parameter $\lambda_1$ controlling sparsity within groups and $\lambda_2$ promoting group-level sparsity.

\subsection{Approximate Message Passing for Sparse Vectors in $\mathbb{R}^N$}
\label{subsec:AMP}
We turn to \ac{AMP} to design an iterative solution to \eqref{eq:SGL}.
The \ac{AMP} framework has been used to recover high-dimensional sparse signals from noisy linear measurements.
In its original form~\cite{donoho2009message}, AMP reconstructs a real-valued sparse vector $\betav \in \mathbb{R}^N$ from observation $\yv$.
At iteration $t$, an effective observation is formed
\begin{align}
\label{eq:effectiveObs}
\rv^{(t)} = \betav^{(t)} + \Xm^{T} \zv^{(t)}
\end{align}
where $\betav^{(t)}$ is the current estimate of the sparse vector, $\Xm^T$ denotes the transpose of the sensing matrix, and $\zv^{(t)}$ is the residual vector described below.
At every step, a refined estimate is obtained via element-wise denoising,
\begin{align}
\label{eq:sparseVectorUpdateAMP}
\betav^{(t+1)} = \eta(\rv^{(t)}),
\end{align}
with soft-thresholding $\eta(\rv^{(t)})
%= \eta(\rv^{(t)}, \lambda^{(t)})
= \prox_{\lambda^{(t)} \|\cdot\|_1}(\rv^{(t)})$ as the proximal operator of the $\ell_1$ norm.
% To reduce correlations between iterations, the residual update incorporates the Onsager correction 
% \vish{Maybe it is better to write the below equations using $t+1$ instead of $t$, i.e.,
% \begin{align}
% \label{eq:residualUpdateAMP}
% \zv^{(t)} = \yv - \Xm \betav^{(t)} + \frac{1}{\delta} \zv^{(t-1)} \langle \eta'(\rv^{(t-1)}) \rangle,
% \end{align}
% }
%
A key feature of the AMP framework is the inclusion of an Onsager correction in the residual computation,
\begin{align}
\label{eq:residualUpdateAMP}
\zv^{(t+1)} = \yv - \Xm \betav^{(t+1)} + \frac{1}{\delta} \zv^{(t)} \langle \eta'(\rv^{(t)}) \rangle,
\end{align}
where $\delta = n/N$ is the measurement ratio and
\begin{equation*}
\textstyle
\langle \eta'(\rv^{(t)}) \rangle \triangleq \frac{1}{N}\sum_{i=1}^N \allowbreak \frac{\partial \eta_i(\rv^{(t)})}{\partial \rv^{(t)}_i}
\end{equation*}
is the  divergence.
The latter term helps preserve approximate independence between the estimates and the residuals, which is crucial to derive key asymptotic properties~\cite{bayati2011dynamics}.
The algorithm is initialized with $\betav^{(0)} = \mathbf{0}$ and $\zv^{(0)} = \yv$.
%Its extension to complex-valued signals was proposed in~\cite{Maleki2013CAMP}.

\section{AMP for Complex Sparse Group LASSO}
\label{sec:AMP_CSGL}

Below, we introduce the \ac{CSGL}-\ac{AMP} algorithm, extending \ac{AMP} to handle \ac{SGL} in complex-valued settings.
This algorithm addresses structured sparse signals with both group-wise and within-group sparsity through a novel complex-valued denoiser with adaptive thresholds.
% This is especially suited for multiple preamble detection in \ac{OTFS}-based random access.
While an initial attempt to solve the \ac{SGL} problem using \ac{AMP} in the real-valued case was presented in~\cite{Chen2022SparseGroupLassoAMP}, to the best of our knowledge, this work constitutes the first effort to apply such an approach to the complex domain.

\subsection{CSGL-AMP Iterative Framework}

Consider the complex linear model in~\eqref{equation:ReceivedSignalRaw}, with sparse vector $\betav \in \mathbb{C}^N$ partitioned into groups.
Only some groups are active, and there are few non-zeros entries within each active group.
In a manner akin to the approach described in Sec.~\ref{subsec:AMP}, \ac{CSGL}-\ac{AMP} estimates $\betav$ using a composite iterative algorithm,
\begin{align}
\label{eq:CSGL_AMP_update}
\rv^{(t)} &= \betav^{(t)} + \Xm^{H} \zv^{(t)} \\
\betav^{(t+1)} &= \eta(\rv^{(t)}, \lambda_1^{(t)}, \lambda_2^{(t)}) \\
\zv^{(t+1)} &= \yv - \Xm \betav^{(t+1)} + \frac{1}{\delta} \zv^{(t)} \left\langle \eta'(\rv^{(t)}, \lambda_1^{(t)}, \lambda_2^{(t)}) \right\rangle
% \zv^{(t)} &= \yv - \Xm \betav^{(t)} + \frac{1}{\delta} \zv^{(t-1)} \left\langle \eta'(\rv^{(t-1)}, \lambda_1^{(t-1)}, \lambda_2^{(t-1)}) \right\rangle,
\end{align}
% \vish{
% \begin{align}
%     \zv^{(t+1)} &= \yv - \Xm \betav^{(t+1)} + \frac{1}{\delta} \zv^{(t)} \left\langle \eta'(\rv^{(t)}, \lambda_1^{(t)}, \lambda_2^{(t)}) \right\rangle,
% \end{align}
% }
%
where $\Xm^{H}$ denotes the Hermitian (conjugate transpose) of the complex-valued sensing matrix, and $\eta(\cdot, \lambda_1, \lambda_2)$ is a complex-valued denoiser tailored to the \ac{CSGL} problem.
Thresholds $\lambda_1$ and $\lambda_2$ are updated adaptively at every iteration through suitable online approximations, allowing the denoiser to dynamically balance sparsity both within and across groups.
The average divergence term appearing in the Onsager correction is obtained by evaluating the Wirtinger derivative of the denoiser $\eta$ \cite{wirtinger1927formalen}, whose derivation is detailed below.

\subsection{CSGL Denoiser}
\label{subsec:CSGL_Denoiser}

Denoiser $\eta(\cdot, \lambda_1, \lambda_2)$ is the proximal operator of the complex-valued \ac{SGL} regularization in \eqref{eq:SGL}.
It is designed to promote double sparsity, as discussed in Sec.~\ref{subsec:StructuredSparsity}. 
Let $\rv \in \mathbb{C}^N$ denote the input vector, partitioned into $G$ disjoint groups as
% \begin{equation}
% \label{eq:groupPartition}
$\mathbf{r} = \begin{bmatrix} \mathbf{r}_1^T & \mathbf{r}_2^T & \cdots & \mathbf{r}_G^T \end{bmatrix}^T$,
% \end{equation}
where $\mathbf{r}_g \in \mathbb{C}^{p_g}$ and \(\sum_{g=1}^G p_g = N\).
Here, $p_g$ denotes the number of elements in the $g$-th group.
Parameters $\lambda_1 > 0$ and $\lambda_2 > 0$ regulate the levels of sparsity within individual elements and across groups, respectively.
The denoising process is structured into two sequential stages.

\subsubsection{Element-wise soft thresholding}
\label{subsubsec:ElementwiseSoftThresholding}
Each coefficient is independently processed using the complex soft-thresholding operator. 
For group $\rv_g$, this yields a thresholded vector
\begin{equation} 
\label{eq:soft_vec}
\sv_g = \eta_{\mathrm{st}}(\rv_g, \lambda_1)
\end{equation}
where the operator is applied element-wise as
\begin{equation} 
\label{eq:soft_scalar}
\eta_{\mathrm{st}}(r, \lambda_1) = 
\begin{cases}
\left(1 - \frac{\lambda_1}{|r|}\right) r & |r| > \lambda_1 \\
0 & \text{otherwise}
\end{cases}
\end{equation}
for each $r \in \rv_g$. This step enforces sparsity within each group by shrinking small-magnitude coefficients toward zero.

\subsubsection{Group-wise shrinkage}
\label{subsubsec:ElementwiseSoftThresholding}

Next, the resulting vector $\sv_g$ is scaled based on its $\ell_2$ norm to enforce group sparsity:
\begin{equation} 
\label{eq:CSGL_Denoiser}
[\eta(\mathbf{r}_g, \lambda_1, \lambda_2)]_j = 
\begin{cases}
 \left(1 - \frac{\lambda_2}{\|\mathbf{s}_g\|_2}\right) \cdot [\sv_g]_j & \text{if } \|\mathbf{s}_g\|_2 > \lambda_2 \\
0 & \text{otherwise}
\end{cases}
\end{equation}
for \(j = 1, \dots, p_g\), where $[\vv]_j$ denotes the $j$-th component of vector $\vv$. 
This step suppresses entire groups whose energy (after element-wise thresholding) is below the $\lambda_2$ threshold.
An equivalent and compact expression for the overall denoiser, acting on each coefficient $r_{g,j}$ in group $\rv_g$, is 
\begin{equation}
\label{eq:CompactCSGLDenoiser}
\begin{split}
\eta_{g,j} \left( \rv \right)
% &= \left( \frac{ \sqrt{ \sum_{\iota=1}^{p_g} \left( |r_{g,\iota}| - \lambda_1 \right)_{+}^2 } - \lambda_2}
% { \sqrt{ \sum_{\iota=1}^{p_g} \left( |r_{g,\iota}| - \lambda_1 \right)_{+}^2 } } \right)
% \left( 1 - \frac{ \lambda_1 }{|r_{g,j}|} \right) r_{g,j} \\
&= \left( 1 - \frac{ \lambda_2 }
{ \sqrt{ \sum_{\iota=1}^{p_g} \left( |r_{g,\iota}| - \lambda_1 \right)_{+}^2 } } \right)
\left( 1 - \frac{ \lambda_1 }{|r_{g,j}|} \right) r_{g,j},
\end{split}
\end{equation}
where $(\cdot)_+ = \max(\cdot, 0)$.
It explicitly combines the element-wise and group-wise effects into a single scaling expression.

\begin{remark}
Special cases: $\lambda_2=0$ yields standard complex \ac{LASSO} (element-wise sparsity only); $\lambda_1=0$ gives \ac{GL} (group sparsity only). Tuning $\lambda_1, \lambda_2$ allows the flexible enforcement of both sparsity levels.
\end{remark}

\subsection{Derivation of the Onsager Correction}
\label{subsec:OnsagerDerivation}

In \ac{AMP}, the Onsager correction accounts for correlations introduced by iterative denoising. For complex-valued denoisers, it can be derived via Wirtinger calculus, treating a complex variable and its conjugate as independent. Let $\eta_{g,j}(\rv)$ denote the $j$-th component of the CSGL denoiser for group $g$, with $\rv$ partitioned as described above.
Its divergence, needed for the Onsager term, is the average of Wirtinger derivatives
\begin{align}
\label{eq:WirtingerDivergence}
\eta_{g,j}'(\rv) = \frac{1}{2} \Big( \frac{\partial \eta_{g,j}}{\partial \Re r_{g,j}} - i \frac{\partial \eta_{g,j}}{\partial \Im r_{g,j}} \Big).
\end{align}
Using the derivative of the modulus
\begin{equation}
\frac{\partial |r_{g,j}|}{\partial \Re r_{g,j}} = \frac{\Re r_{g,j}}{|r_{g,j}|}, \quad
\frac{\partial |r_{g,j}|}{\partial \Im r_{g,j}} = \frac{\Im r_{g,j}}{|r_{g,j}|},
\end{equation}
and the chain rule on~\eqref{eq:CompactCSGLDenoiser}, one obtains $\partial \eta_{g,j}/\partial \Re r_{g,j}$; a similar expression holds for $\partial \eta_{g,j}/\partial \Im r_{g,j}$. Substituting into~\eqref{eq:WirtingerDivergence} gives the complex divergence for each nonzero element.

The global Onsager term averages over all components
\begin{equation}
\label{eq:OnsagerAverage}
\textstyle
\frac{1}{N} \sum_{g} \sum_{j \in \mathcal{J}_g} \eta_{g,j}'(\rv),
\end{equation}
where $\mathcal{J}_g$ is the set of nonzero indices in group $g$. After simplification, the closed-form expression reads
\begin{equation*}
\begin{split}
\label{eq:OnsagerFinal}
\frac{1}{N} \! \sum_{g \in \mathcal{G}_a} \! \Bigg(
|\mathcal{J}_g| - \sum_{j \in \mathcal{J}_g} \frac{\lambda_1}{2 |r_{g,j}|} 
- \frac{\lambda_2}{2} \frac{2|\mathcal{J}_g| - 1 - \sum_{j \in \mathcal{J}_g} \frac{\lambda_1}{2 |r_{g,j}|}}{\sqrt{\sum_\iota (|r_{g,\iota}|-\lambda_1)_+^2}}
\Bigg)
\end{split}
\end{equation*}
where $\mathcal{G}_a$ is the set of active groups ($\|\sv_g\|_2>\lambda_2$).

This expression captures the interplay between $\lambda_1$, $\lambda_2$, and the residual geometry. $|\mathcal{J}_g|$ reflects the group size, while other terms adjust for thresholded magnitudes. When $\lambda_2=0$, it reduces to the complex soft-threshold divergence (standard LASSO), and when $\lambda_1=0$, it reflects pure group-wise shrinkage, consistent with the CSGL structure and the two-stage denoising process in Sec.~\ref{subsec:CSGL_Denoiser}.

\section{Numerical Results}
\label{sec:NumRes}

% \subsection{Simulation Setup}
% \label{subsec:SimSetup}

We conducted Monte Carlo simulations to evaluate the proposed \ac{CSGL}-\ac{AMP} algorithm against the \ac{OST} method~\cite{mattu2024delay}.
While group-sparse compressed sensing has been studied~\cite{dong2025group}, existing works do not address the two-level sparsity of our model.
Hence, we restrict the comparison to \ac{OST}, which is tailored for multiple preamble detection in \ac{OTFS}-based systems.
The simulations use an \ac{OTFS} sensing matrix built from the \ac{Veh-A} channel model as in~\cite{mattu2024delay}.
The matrix $\Xm \in \mathbb{C}^{n \times N}$ has $n = M_\mathrm{dd} N_\mathrm{dd}$ rows, corresponding to the length of each preamble, and $N = G \cdot |\mathcal{S}|$ columns.
% , where $G$ denotes the number of employed preambles (e.g., \ac{OTFS}-modulated \ac{ZC} sequences) and $\mathcal{S}$ represents the set of candidate \ac{DD} shifts.
Each preamble $\sv_j \in \mathbb{C}^n$ is shifted across all $(\tau, \nu) \in \mathcal{S}$ to form a group of $|\mathcal{S}|$ columns in $\Xm$. We set $n=1147$ with $M_\mathrm{dd}=31$, $N_\mathrm{dd}=37$, $|\mathcal{S}|=20$, and we assume a collision-free setting where each active user selects a distinct preamble.
The unknown sparse vector $\betav \in \mathbb{C}^{G|\mathcal{S}|}$ exhibits:
(\textit{i}) group sparsity because $K$ of the $G$ preambles are active;
(\textit{ii}) within-group sparsity, as each active preamble is shaped by a multipath channel with $L=6$ dominant $(\tau,\nu)$ components, each with a random complex phase.
Performance is measured by the misdetection probability $P_\mathrm{md}$, the ratio of undetected to transmitted preambles. We first compare $P_\mathrm{md}$ versus \ac{SNR} for fixed $G$ and $K$, which define the measurement ratio $\delta = n/N$ and group sparsity ratio $\rho_\mathrm{G} = K/G$. Then, by varying $K$ and $G$, we explore the limits of each method in achieving a target $P_\mathrm{md}$ across a broad range of sparsity and measurement conditions.

% \subsection{Misdetection Probability as a Function of SNR}
% %
% \begin{figure}
%     \centering
%     \resizebox{0.99\columnwidth}{!}{
%         \input{Figures/Plot_AMPvsOST_OTFS_VarSNR}
%     }
%     \caption{Probability of misdetection $P_\mathrm{md}$ versus \ac{SNR} for the proposed \ac{CSGL}-\ac{AMP} algorithm and the\ac{OST} method from \cite{mattu2024delay}.}
%     \label{fig:Plot_AMPvsOST_OTFS_VarSNR}
% \end{figure}
% %
%
\begin{figure}
    \centering
    \resizebox{0.93\columnwidth}{!}{
\begin{tikzpicture}

\begin{axis}[%
width=4.521in,
height=3.566in,
at={(0.758in,0.481in)},
scale only axis,
xmin=4,
xmax=18,
xlabel style={font=\color{white!15!black}, font = \large},
xlabel={SNR [dB]},
ymode=log,
ymin=0.0001,
ymax=1,
yminorticks=true,
ylabel style={font=\color{white!15!black}, font = \large},
ylabel={$P_\mathrm{md}$},
ytick = {1e-4, 1e-3, 1e-2, 1e-1, 1},
yticklabels = {$10^{-4}$, $10^{-3}$, $10^{-2}$, $10^{-1}$, $1$},
tick label style={black, semithick, font=\large},
xmajorgrids,
ymajorgrids,
axis background/.style={fill=white},
legend style={legend cell align=left, align=left, draw=white!15!black, font = \large, legend pos = south west}
]

\addplot [color=red, line width = 1.5pt,mark=square, mark size = 3pt, mark options={solid, red}]
  table[row sep=crcr]{%
0	0.546315789473684\\
2	0.424736842105263\\
4	0.33078947368421\\
6	0.223684210526316\\
8	0.153684210526316\\
10	0.102105263157895\\
12	0.0805263157894737\\
14	0.0663157894736843\\
16	0.0562793121417404\\
18	0.0463659147869674\\
20	0.0445154943433349\\
};
\addlegendentry{OST}

\addplot [color=blue, dashed, line width = 1.5pt, mark=square, mark size = 3pt, mark options={solid, blue}]
table[row sep=crcr]{%
0   0.777105263157895\\
2   0.775526315789473\\
4   0.746315789473685\\
6   0.697105263157895\\
8   0.650789473684210\\
10  0.570263157894737\\
12  0.479210526315790\\
14  0.376578947368421\\
16  0.296578947368421\\
18  0.228421052631579\\
};
\addlegendentry{CL-AMP}

\addplot [color=blue, dash dot, line width = 1.5pt, mark=square, mark size = 3pt, mark options={solid, blue}]
  table[row sep=crcr]{%
0   0.704736842105263\\	
2   0.642368421052631\\	
4   0.560263157894737\\
6   0.475526315789474\\
8   0.370526315789474\\
10  0.259473684210526\\
12  0.172894736842105\\	
14  0.100263157894737\\	
16  0.0668214654282766\\
18  0.0417259364627785\\
};
\addlegendentry{CGL-AMP}

\addplot [color=blue, line width = 1.5pt, mark=square, mark size = 3pt, mark options={solid, blue}]
  table[row sep=crcr]{%
0   0.767894736842105\\	
2   0.700263157894737\\
4   0.591052631578947\\
6   0.418684210526316\\
8   0.227368421052632\\
10  0.0719298245614035\\
12  0.0176925364258103\\
14  0.00302033492822965\\
16  0.000250000000000000\\	
%18   1.31578947368421e-05\\
};
\addlegendentry{CSGL-AMP}

\end{axis}
\end{tikzpicture}%
    }
    \caption{Probability of misdetection $P_\mathrm{md}$ versus \ac{SNR} for the proposed \ac{CSGL}-\ac{AMP} algorithm and the \ac{OST} algorithm from \cite{mattu2024delay}. The block sparsity is set to $\rho_\mathrm{G} = 0.2$, and the measurement ratio is $\delta = 0.3$.}
    \label{fig:Plot_AMPvsOST_OTFS_VarSNR_rho0_2_delta0_3}
\end{figure}
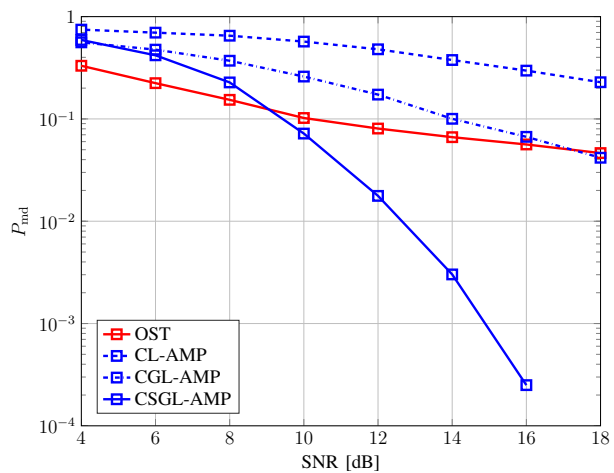
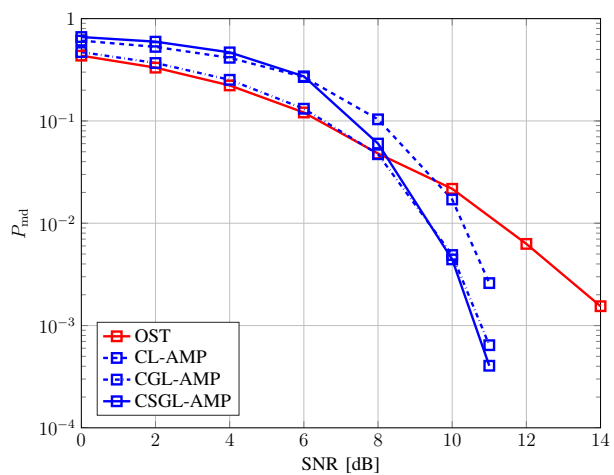
\begin{figure}
    \centering
    \resizebox{0.93\columnwidth}{!}{
\begin{tikzpicture}

\begin{axis}[%
width=4.521in,
height=3.566in,
at={(0.758in,0.481in)},
scale only axis,
xmin=0,
xmax=14,
xlabel style={font=\color{white!15!black}, font = \large},
xlabel={SNR [dB]},
ymode=log,
ymin=0.0001,
ymax=1,
yminorticks=true,
ylabel style={font=\color{white!15!black}, font = \large},
ylabel={$P_\mathrm{md}$},
ytick = {1e-4, 1e-3, 1e-2, 1e-1, 1},
yticklabels = {$10^{-4}$, $10^{-3}$, $10^{-2}$, $10^{-1}$, $1$},
tick label style={black, semithick, font=\large},
xmajorgrids,
ymajorgrids,
axis background/.style={fill=white},
legend style={legend cell align=left, align=left, draw=white!15!black, font = \large, legend pos = south west}
]
% \addplot [color=red, line width = 1.2pt, mark=o, mark size = 3pt, mark options={solid, red}]
%   table[row sep=crcr]{%
% 0	0.537\\
% 2	0.4105\\
% 4	0.2605\\
% 6	0.154901960784314\\
% 8	0.0662393162393162\\
% 10	0.0271226415094339\\
% 12	0.00766917293233081\\
% 14	0.00297441998810231\\
% %16	0.00111061750333185\\
% 18	0\\
% 20	0\\
% };
% \addlegendentry{OST - $\rho_\mathrm{B}= 0.175, \delta = 0.5$}

\addplot [color=red, line width = 1.5pt, mark=square, mark size = 3pt, mark options={solid, red}]
  table[row sep=crcr]{%
0	0.434761904761905\\
2	0.331904761904762\\
4	0.222380952380952\\
6	0.120952380952381\\
8	0.0479853479853479\\
10	0.0217174736003188\\
12	0.00628219625581102\\
14	0.00154557116582433\\
%16	0.00056563085516315\\
18	0\\
20	0\\
};
\addlegendentry{OST}

\addplot [color=blue, dashed, line width = 1.5pt, mark=square, mark size = 3pt, mark options={solid, blue}]
  table[row sep=crcr]{%
% 0   0.605714285714286\\	
% 2   0.512857142857143\\	
% 4   0.411904761904762\\
% 6   0.266666666666666\\
% 8   0.0990476190476190\\	
% 10  0.0158730158730159\\
0   0.609523809523810\\
2   0.529523809523810\\
4   0.414285714285714\\
6   0.268571428571428\\
8   0.103693813974188\\	
10  0.0170739348370927\\
11  0.002599742599743\\
};
\addlegendentry{CL-AMP}

\addplot [color=blue, dash dot, line width = 1.5pt, mark=square, mark size = 3pt, mark options={solid, blue}]
  table[row sep=crcr]{%
0   0.470952380952381\\	
2   0.368571428571428\\
4   0.252380952380952\\
6   0.131652661064426\\
8   0.0472739820565907\\
10  0.00489269970450031\\
11  6.428571428571436e-04\\
};
\addlegendentry{CGL-AMP}

\addplot [color=blue, line width = 1.5pt, mark=square, mark size = 3pt, mark options={solid, blue}]
  table[row sep=crcr]{%
% 0	0.667619047619048\\
% 2	0.61\\
% 4	0.475714285714286\\
% 6	0.255238095238095\\
% 8	0.0642857142857142\\
% 10	0.003494975972040\\
% 11  3.195909236177696e-04\\
0   0.660952380952381\\	
2   0.595714285714286\\
4   0.468095238095238\\
6   0.272857142857142\\
8   0.0600907029478457\\
10  0.00443169968717413\\	
11  0.000404761904761905\\
};
\addlegendentry{CSGL-AMP}

\end{axis}
\end{tikzpicture}%
    }
    \caption{Probability of misdetection $P_\mathrm{md}$ versus \ac{SNR} for the proposed \ac{CSGL}-\ac{AMP} algorithm and the \ac{OST} algorithm from \cite{mattu2024delay}. The block sparsity is set to $\rho_\mathrm{G} = 0.3$, and the measurement ratio is $\delta = 0.8$.}
    \label{fig:Plot_AMPvsOST_OTFS_VarSNR_rho0_3_delta0_8}
\end{figure}
In Fig.\ref{fig:Plot_AMPvsOST_OTFS_VarSNR_rho0_2_delta0_3} and Fig.\ref{fig:Plot_AMPvsOST_OTFS_VarSNR_rho0_3_delta0_8}, we plot $P_\mathrm{md}$ versus \ac{SNR} for the proposed \ac{CSGL}-\ac{AMP}, compared with the \ac{OST} method.
To assess the role of two-level sparsity, we also include the \ac{CL}-\ac{AMP} and \ac{CGL}-\ac{AMP} variants.
Fig.\ref{fig:Plot_AMPvsOST_OTFS_VarSNR_rho0_2_delta0_3} considers an underdetermined regime with $\delta = 0.3$, $G = 191$ Zadoff-Chu preambles, and $\rho_\mathrm{G} = 0.2$ ($K=38$).
Here, \ac{CSGL}-\ac{AMP} substantially outperforms the alternatives, achieving the lowest $P_\mathrm{md}$ in the practical region of interest ($P_\mathrm{md}\leq10^{-1}$).
Notably, \ac{OST} shows a performance floor and cannot reach $P_\mathrm{md}=10^{-2}$ even at high \ac{SNR}, whereas \ac{CSGL}-\ac{AMP} attains this at an \ac{SNR} of approximately $12.5~\mathrm{dB}$.
The weaker results of \ac{CL}-\ac{AMP} and \ac{CGL}-\ac{AMP} highlight the need to exploit both group and within-group sparsity.

Fig.\ref{fig:Plot_AMPvsOST_OTFS_VarSNR_rho0_3_delta0_8} depicts a less sparse case with $\delta=0.8$, $G=72$, and $\rho_\mathrm{G}=0.3$ ($K=21$).
The performance gap narrows, yet \ac{CSGL}-\ac{AMP} still leads in the low $P_\mathrm{md}$ regime.
For instance, achieving $P_\mathrm{md}=10^{-2}$ requires only $9~\mathrm{dB}$, a $2~\mathrm{dB}$ gain over \ac{OST} ($11~\mathrm{dB}$). At lower \ac{SNR}, \ac{OST} can outperform, but with proper tuning, \ac{CGL}-\ac{AMP} approaches its performance, underscoring the flexibility of the proposed framework across different operating regimes.

% \subsection{Effect of Block Sparsity and Measurement Ratio}
In Fig.~\ref{fig:Plot_AMPvsOST_OTFS_VarSparsityVarMeas}, we compare \ac{CSGL}-\ac{AMP} with \ac{OST} in terms of ``validity regions'', i.e., the delta $(\delta,\rho_\mathrm{G})$ region where a target misdetection probability $P_\mathrm{md}^\star = 10^{-2}$ is met. Each curve marks the upper boundary of the region: all points below satisfy the target. Results are shown for two \ac{SNR} levels.
At $10~\mathrm{dB}$, \ac{CSGL}-\ac{AMP} exhibits a larger validity region than \ac{OST}, although both are ultimately limited by noise at high~$\rho_G$.
At $20~\mathrm{dB}$, \ac{CSGL}-\ac{AMP} remains superior, with higher \ac{SNR} enlarging the region to support more active users.
Although \ac{AMP} requires multiple iterations, each step involves only simple matrix-vector multiplications, making the per-iteration complexity comparable to \ac{OST} and scalable to large dimensions.

\begin{figure}
    \centering
    \resizebox{0.887\columnwidth}{!}{
\begin{tikzpicture}

\begin{axis}[%
width=4.521in,
height=3.566in,
at={(0.758in,0.481in)},
scale only axis,
xmin=0.056,
xmax=0.95,
xlabel style={font=\color{white!15!black}, font = \large},
xlabel={Measurement Ratio $\delta$},
ymin=0,
ymax=1,
ylabel style={font=\color{white!15!black}, font = \large},
ylabel={$\rho_\mathrm{G}$},
axis background/.style={fill=white},
tick label style={black, semithick, font=\large},
xmajorgrids,
ymajorgrids,
legend style={legend cell align=left, align=left, draw=white!15!black, font=\large, legend pos = north west}
]
\addplot [color=red, dashed, line width = 1.7pt, name path=OST10]
  table[row sep=crcr]{%
% 0.3	0.05\\
% 0.4	0.05\\
% 0.5	0.1\\
% 0.6	0.1\\
% 0.7	0.15\\
% 0.8	0.2\\
% 0.9	0.2\\
% 1	0.25\\
0.056 0.008\\
0.1 0.01\\
0.2 0.03\\
0.3	0.05\\
0.475	0.0921875\\
0.65	0.11875\\
0.825	0.2\\
1	0.25\\
};
\addlegendentry{OST - SNR = $10~\mathrm{dB}$}

\addplot [color=red, solid, line width = 1.7pt, name path=OST20]
  table[row sep=crcr]{%
% 0.2	0.05\\
% 0.3	0.1\\
% 0.4	0.15\\
% 0.5	0.25\\
% 0.6	0.3\\
% 0.7	0.35\\
% 0.8	0.45\\
% 0.9	0.55\\
% 1	0.95\\
0.0560 0.02\\
0.1 0.04\\
0.2	0.05\\
0.3	0.1\\
0.475	0.228125\\
0.65	0.322916666666667\\
0.825	0.4721875\\
0.9 0.53\\
0.92 0.55\\
0.95 0.60\\
1	0.95\\
};
\addlegendentry{OST - SNR = $20~\mathrm{dB}$}

\addplot [color=blue, dashed, line width = 1.7pt, name path=AMP10]
  table[row sep=crcr]{%
0.056 0.04\\
0.1 0.04\\
0.2	0.05\\
0.3	0.1\\
0.475	0.2\\
% 0.65	0.3\\
% 0.825	0.35\\
% 1	0.4\\
0.6 0.26\\
0.7 0.288\\
0.8 0.34\\
0.85 0.37\\
0.95 0.39\\   
};
\addlegendentry{CSGL-AMP - SNR = $10~\mathrm{dB}$}

\addplot [color=blue, solid, line width = 1.7pt, name path=AMP20]
  table[row sep=crcr]{%
0.056 0.04\\
0.1	0.08\\
0.2 0.15\\	
0.4	0.3\\
0.6	0.5\\
0.8 0.7\\
0.9 0.84\\
0.95 0.89\\
};
\addlegendentry{CSGL-AMP - SNR = $20~\mathrm{dB}$}

% X-axis (y=0), define as path "axis"
\path[name path=axis] (axis cs:0.056,0) -- (axis cs:0.95,0);

\addplot [red,opacity=0.025] fill between[of=OST10 and axis];
\addplot [red,opacity=0.025] fill between[of=OST20 and axis];
\addplot [blue,opacity=0.025] fill between[of=AMP10 and axis];
\addplot [blue,opacity=0.025] fill between[of=AMP20 and axis];

\end{axis}
\end{tikzpicture}%
    }
    \caption{Validity regions in the $(\rho_\mathrm{G}, \delta)$ plane where the target misdetection probability \(P_\mathrm{md}^{\star} \leq 10^{-2}\) is met, shown for two \ac{SNR} levels. The comparison is carried out between the proposed \ac{CSGL}-\ac{AMP} algorithm and the \ac{OST} algorithm from \cite{mattu2024delay}.}
    \label{fig:Plot_AMPvsOST_OTFS_VarSparsityVarMeas}
\end{figure}
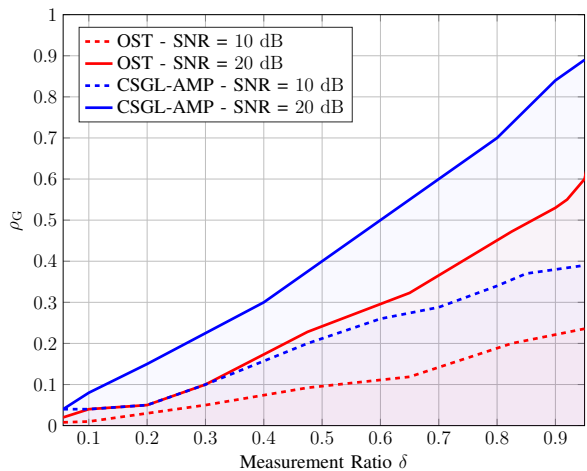

\section{Conclusions}
\label{sec:Conclusions}
We presented a novel \ac{CSGL}-\ac{AMP} algorithm for preamble detection in \ac{OTFS}-based random access, which judiciously exploits both group and within-group sparsity to achieve performance gains over existing methods, as shown in simulations under the \ac{Veh-A} channel model.
Future work will extend the framework to collision scenarios and fully grant-free access in the \ac{DD} domain, with joint preamble detection, channel estimation, and data recovery, further demonstrating the potential of structured \ac{AMP} methods to enhance reliability and scalability in massive machine-type communications.
% We presented a novel \ac{CSGL}-\ac{AMP} algorithm for preamble detection in \ac{OTFS}-based random access.
% By judiciously exploiting both group and within-group sparsity, the new framework offers performance improvements over existing methods, as shown in simulations under the \ac{Veh-A} channel model.
% Future work will extend the framework to collision scenarios and full grant-free random access in the \ac{DD} domain, where preamble detection, channel estimation, and data recovery are jointly addressed.
% This direction will further demonstrate the potential of structured \ac{AMP} methods to enhance reliability and scalability in massive machine-type communication systems.

% \section{Funding Acknowledgements}
% University of Bologna team is supported by the European Union -- Next Generation EU under the Italian National Recovery and Resilience Plan (NRRP), partnership on ``Telecommunications of the Future'' (PE00000001 -- program ``RESTART'').
% This material is based, in part, upon work conducted at Texas A\&M University and supported by the National Science Foundation (NSF) under Grant CNS-2148354.

% \section{Compliance with Ethical Standards}
% This is a numerical simulation study for which no ethical approval was required.

%%%%%%%%%%%%%%%%%%%%%%%%%%%%%%%%%%%%%%%%%%%%%%%%%%%%%%
%%%%%%%%%%%%%%%%%%%%%%%%%%%%%%%%%%%%%%%%%%%%%%%%%%%%%%
\bibliographystyle{IEEEtran}
\bibliography{IEEEabrv, refs}
%%%%%%%%%%%%%%%%%%%%%%%%%%%%%%%%%%%%%%%%%%%%%%%%%%%%%%
%%%%%%%%%%%%%%%%%%%%%%%%%%%%%%%%%%%%%%%%%%%%%%%%%%%%%%

\end{document}